\documentclass[prl,twocolumn,floatfix,superscriptaddress,showpacs]{revtex4}%
\usepackage{graphicx}
\usepackage{bm}

\begin{document}
\title{Transverse Instability of Avalanches in Granular Flows down Incline}

\author{Igor S.~Aranson}
\affiliation{Materials Science Division, Argonne National
Laboratory, 9700 South Cass Avenue, Argonne, IL 60439}
\author{Florent Malloggi}
\affiliation{Laboratoire de Physique et M\'ecanique des Milieux
H\'et\'erog\`enes, 10 rue Vauquelin 75005 Paris France, UMR CNRS
7636}
\author{Eric Cl\'ement}
\affiliation{Laboratoire de Physique et M\'ecanique des Milieux
H\'et\'erog\`enes, 10 rue Vauquelin 75005 Paris France, UMR CNRS
7636}

\date{\today}

\begin{abstract}
Avalanche experiments on an erodible substrate are treated in the
framework of ``partial fluidization'' model of dense granular
flows. The model identifies a family of propagating soliton-like
avalanches with shape and velocity controlled by the inclination
angle and the depth of  substrate. At high inclination angles the
solitons display a transverse instability, followed by coarsening
and fingering  similar to recent experimental observation. A
primary cause for the transverse instability is directly related
to the dependence of soliton velocity on the granular mass trapped
in the avalanche.
\end{abstract}
\pacs{47.10.+g,  68.08.-p, 68.08.Bc}

\maketitle

Granular deposit instabilities are ubiquitous in nature; they
display solid or fluid-like behavior as well as catastrophic
events such as avalanches, mud flows or land slides. A somewhat
similar phenomena unfold below sea level. Their occurrence is
relevant for a broad variety of marine-based technologies, such as
off-shore oil exploitation or deep-sea telecommunication cables,
and is a matter of concern for coastal communities. The
perspective of risk modelling of these unstable matter waves is
hindered by the lack of conceptual clarity since the conditions
triggering avalanches and the rheology of the particulate flows
are poorly understood. While extensive laboratory-scale
experiments on dry and submerged granular materials flowing on
rough inclined plane
\cite{GDR04,Pouliquen05,Daerr:1999,Malloggi:2005,Borzsonyi:2005,Pouliquen:1997}
have brought new perspectives for the elaboration of reliable
constitutive relations, many open questions still remain such as
avalanches propagation on erodible substrates. It has been shown
experimentally that families of localized unstable avalanche waves
can be triggered in the bi-stability  domain of phase diagram
\cite{Daerr:1999}. Also, the shape of localized droplet-like waves
was recently shown to depend strongly on the intimate nature of
the granular material used \cite{Borzsonyi:2005}. All these
questions are closely related to the compelling need for a
reliable description of the fluid/solid transition for particulate
assemblies in the vicinity of flow arrest. Recent avalanche
experiments on erodible layers performed both in air and under
water\cite{Malloggi:2005} though strongly differing by spatial and
time scales involved, display striking common features: solitary
quasi one-dimensional waves transversally unstable at higher
inclination angles. The instability further develops into a
fingering pattern via a coarsening scenario. So far, this
phenomenology, likely to be common to many natural
erosion/deposition processes, misses a clear physical explanation.
 From a theoretical perspective, a model of
``partially fluidized'' dense granular flows was recently
developed to couple a phenomenological description of a
solid/fluid transition with hydrodynamic transport equations. It
reproduces  many features found experimentally such as
metastability of a granular deposit, triangular down-hill and
balloon-type up-hill avalanches and variety of shear flow
instabilities \cite{at2001,at2005}. The model was later calibrated
with molecular dynamics simulations \cite{VTA03}.

In this Letter the partial fluidization model is applied to
avalanches on a thin erodible sediment layer. A set of equations
describing the dynamics of fully eroding waves is derived, and a
 family of soliton solutions
propagating downhill is obtained.  The velocity and shape
selection of these solitons is investigated as well as the
existence of a linear transverse instability. The primary cause of
the instability is identified with the dependence of soliton
velocity on its trapped mass.  A numerical study is conducted to
follow nonlinear evolution of
 avalanche front. All these features are discussed in the context of the
experimental
  findings of Malloggi et al.\cite{Malloggi:2005}. New perspectives for quantitative contact between
  modelling and experiments are then underlined.

 According to the partial fluidization theory \cite{at2001},
 the ratio of the static part of shear stress  to the fluid part
of the full stress tensor is controlled by an order parameter (OP)
$\rho$, which  is scaled in such a way that in granular solid
$\rho = 1$ and in the fully developed flow (granular liquid) $\rho
\to 0$. At the ``microscopic level'' OP is defined as a fraction
of the number of persistent particle contacts to the total number
of contacts. Due to a strong dissipation in dense granular flows,
$\rho$ is assumed to obey purely relaxational dynamics controlled
by the Ginzburg-Landau equation for generic first order phase
transition,
\begin{equation}
\tau_{\rho} \frac{D\rho}{Dt}=l_{\rho}^2
\nabla^2\rho-\frac{\partial {F}(\rho, \delta)}{\partial \rho}.
\label{GL2}
\end{equation}
Here $\tau_{\rho},l_{\rho}\approx d $ are the OP characteristic
time and length scales, $d$ is the grain size. ${F}(\rho, \delta)$
is a free energy density which is postulated to have two local
minima at $\rho=1$ (solid phase) and $\rho=0$ (fluid phase) to
account for the bistability near the solid-fluid transition. The
relative stability of the two phases is controlled by the
 parameter $\delta$ which in turn is determined
by the stress tensor.  The simplest assumption consistent with the
Mohr-Coulomb yield criterion is to take it as a function of
$\phi=\max |\sigma_{mn}/\sigma_{nn}|$, where the maximum is sought
over all possible orthogonal directions $m$ and $n$.

For  thin  layers on inclined plane  Eq. (\ref{GL2}) can be
simplified by fixing the structure of OP in $z$-direction ($z$
perpendicular to the bottom, $x$ is directed down the chute and
$y$ in the vorticity direction) $\rho=1-A(x,y) \sin(\pi z/2 h)$,
$h$ is the local layer thickness, $A$ is slowly-varying function.
This approximation valid for  thin layers when there is no
formation of static layer beneath  the avalanche. Then one obtains
equations governing the evolution $h$ and $A$, coordinates $x,y$,
height $h$,  and time $t$ are normalized by $l_{\rho},\tau_{\rho}$
correspondingly \cite{at2001,at2005},
\begin{eqnarray}
\frac{\partial h}{\partial t} &=&  - \alpha \frac{\partial h^3
A}{\partial x}+ \frac{\alpha}{\phi}  \nabla \left(h^3 A \nabla h
\right)
\label{conser} \\
\frac{\partial A} {\partial t}  &=& \lambda_0 A+ \nabla^2 A
+\frac{8(2-\delta) }{3 \pi} A^2 -\frac{3 }{4}  A^3 \label{A1}
\end{eqnarray}
where $\nabla^2= \partial_x^2+\partial_y^2$,
$\lambda_0=\delta-1-\pi^2/4 h^2$,  dimensionless transport
coefficient:
\begin{equation}
\alpha \approx \frac{ 2 (\pi^2-8)}{\pi^3 \mu}  g \tau_\rho l_\rho
\sin \bar \varphi, \label{alpha}
\end{equation}
$\mu$ is the shear viscosity, $\bar \varphi$ is the chute
inclination, $\phi=\tan \bar \varphi$. Control parameter $\delta$
includes a correction due to the change in the local slope
$\delta=\delta_0+\beta h_x$, $\beta \approx 1.5-3$ depending on
the value of $\bar \varphi$, see for detail \cite{at2001,at2005}.
The last term in Eq. (\ref{conser}) is also due to change of local
slope and is obtained from expansion $\varphi=\bar \varphi + h_x$.
This term is responsible for the saturation of the slope of the
avalanche front (without it the front can be arbitrary steep)
\cite{at2005}.

\begin{figure}[t!]
\includegraphics[width=3.4in]{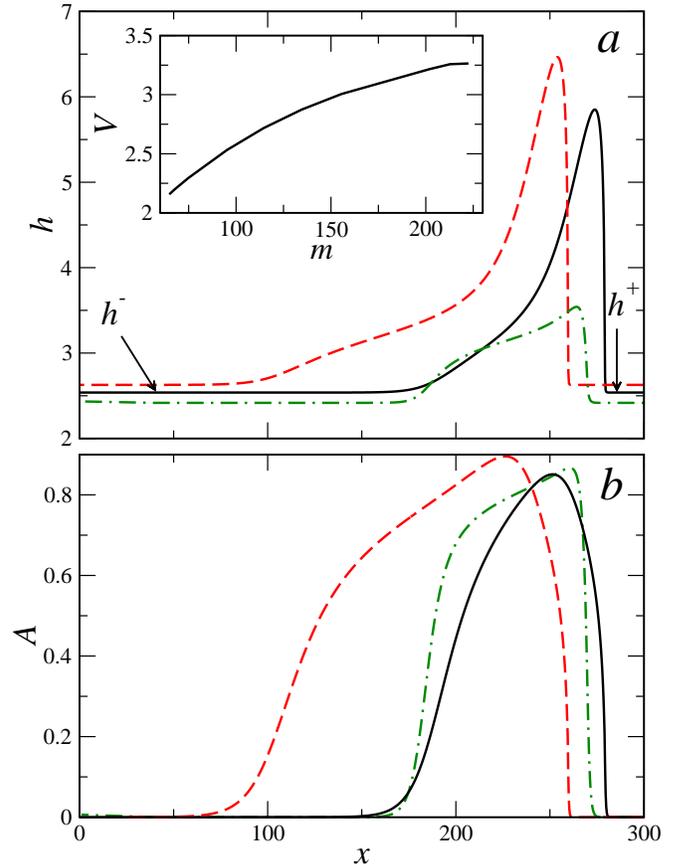}
\caption{ $h$ (a) and $A$ (b)   for various values of $m$ and
$\alpha$. Solid line is for $m=147.7$, $V=2.72$, dashed line is
for $m=211$, $V=3.12$, for $\delta=1, \alpha=0.08, \beta=2$;
point-dashed line is for $\alpha=0.025,\delta=1.15$, $m=62$,
$V=0.86$. Inset: $V$ vs $m$.} \label{fig1}
\end{figure}

In the coordinate system co-moving with the velocity $V$ Eqs.
(\ref{conser}),(\ref{A1}) assume the form:
\begin{eqnarray}
\frac{\partial h}{\partial t} &=& V \partial_x h  - \alpha
\frac{\partial h^3 A}{\partial x}+ \frac{\alpha}{\phi}  \nabla
\left(h^3 A \nabla h \right)
\label{conser11} \\
\frac{\partial A} {\partial t}  &=& V \partial_x A+  \lambda_0 A+
\nabla^2 A +\frac{8(2-\delta) }{3 \pi} A^2 -\frac{3 }{4}  A^3
\label{A_11}
\end{eqnarray}
Numerical studies revealed that the one-dimensional Eqs.
(\ref{conser11}),(\ref{A_11}) possess a one-parametric family of
localized (solitons) solutions, see Fig \ref{fig1}:
\begin{equation}
A(x,t)=A(x-Vt), h(x,t)=h(x-Vt) \label{soliton}
\end{equation}
Here the boundary conditions take a form $ h \to h_0,A \to 0 $ for
$ x \to \pm \infty$, where $h_0$ is the asymptotic height. The
one-dimensional  steady state soliton solution (\ref{soliton})
satisfy:
\begin{eqnarray}
V (h-h_0)  &=&  \alpha  h^3 A \left( 1 - \frac{\partial_x
h}{\phi}
 \right)
\label{conser1} \\
-V \frac{\partial A} {\partial x}  &=& \lambda A+
\partial_x^2 A +\frac{8(2-\delta) }{3 \pi} A^2 -\frac{3 }{4}
A^3 \label{A11}
\end{eqnarray}

The solutions can be parameterized by the ``trapped mass'' $m$
carried by the soliton, i.e. the area above $h_0$,
\begin{equation}
m=\int_{-\infty}^\infty (h-h_0) dx \label{mass}
\end{equation}
The velocity $V$ is an increasing function of $m$, see inset Fig.
\ref{fig1}a. The structure of the solutions is sensitive to the
value  of $\alpha$: for large $\alpha$ the solution has a
well-pronounced shock-wave shape,  Fig. \ref{fig1}, with the
height of the crest $h_{max}$ several times larger than the
asymptotic depth $h_0$. For $\alpha \to 0$ the solution assumes
more rectangular form, see Fig. \ref{fig1}, and $h_{max}-h_0 \ll
h_0$.

\begin{figure}[t!]
\includegraphics[width=2.15in,angle=-90]{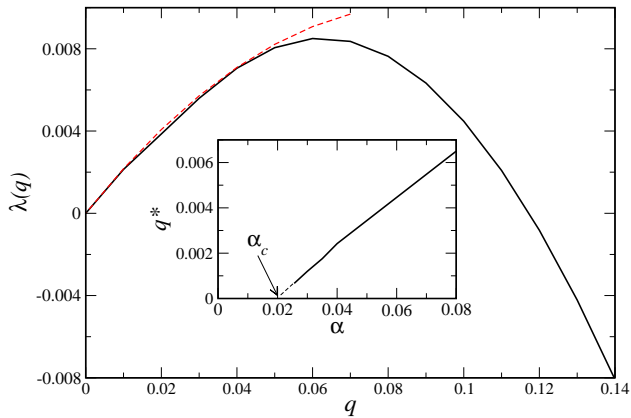}
\caption{$\lambda(q)$ vs $q$ for $\delta=1.15$, $\alpha=0.08$ and
$m=102$. Solid line: $\lambda(q)$ obtained by numerical stability
analysis of one-dimensional solution Eq. (\ref{soliton1}). Dashed
line is solution of Eq. (\ref{l0}). Inset: optimal wavenumber of
$q^*$ vs $\alpha$ for $\delta=1.15$} \label{fig2}
\end{figure}

To understand {\it transverse instability}  we focus on the
soliton solution with slowly varying position $x_0(y,t)$
\begin{eqnarray}
A(x,t)=\bar A(x-x_0(t,y)), \;
 h(x,t)= \bar h(x-x_0(t,y))
\label{soliton1}
\end{eqnarray}
%We also explicitly included that the steady state solutions $\bar
%A, \bar h$ are parameterized by the is mass $m$.
Substituting Eq.
(\ref{soliton1}) in Eq. (\ref{conser11}) and integrating over $x$,
one obtains
\begin{equation}
\partial_t m = V(m) (h^+-h^-(m)) - \zeta_1 \partial_y^2 x_0+ \zeta_2 \partial_y^2 m
\label{mass1}
\end{equation}
where $\zeta_{1,2}=const$ is defined as
\begin{eqnarray*}
\zeta_1 = \frac{ \alpha}{\phi} \int _{-\infty} ^\infty \left(\bar
A \bar h^3
\partial_x
\bar h \right) dx, \;
 \zeta_2 = \frac{ \alpha}{\phi} \int
_{-\infty} ^\infty \left(\bar A \bar h^3
\partial_m \bar h \right)dx
%\label{zeta}
\end{eqnarray*}
%Last term  $\zeta_2 \partial_y^2 m$  in Eq. (\ref{mass1}) is
%formally of higher order and can be neglected.
Here $h^+=h(x\to \infty)$ is the height of the deposit layer ahead
of the front and $h^-=h(x\to -\infty) $ is the height behind the
front, see Fig. \ref{fig1}a. While the value of $h^+$ is
prescribed by the initial sediment height, the value of $h^-$
behind the front is determined by the velocity (or mass) of the
front. For steady-state  solution $h^+=h^-=h_0$. For the
slowly-evolving solution the difference between $h^+$ and $h^-$
can be small, however it is important for the stability analysis.
These terms are also necessary to describe experimentally observed
initial acceleration/slowdown of the avalanches. Substituting Eqs.
(\ref{soliton1}) into Eq. (\ref{A1}) and performing orthogonality
conditions one obtains
\begin{equation}
\partial_t x_0= V(m)  +\partial_y^2 x_0   \label{V1}
\end{equation}
There are also higher order  terms in  Eq. (\ref{V1}) which we
neglect for simplicity. To see the onset of the instability we
keep only the leading terms in Eq.(\ref{mass1}),(\ref{V1}), using
$ V(m)\approx V(m_0)+ V_m (m-m_0) $, and $\tilde m = m-m_0 \ll
m_0$:
\begin{eqnarray}
\partial_t \tilde m &=&- \tau \tilde m  - \zeta_1 \partial_y^2 x_0+ \zeta_2  \partial_y^2 \tilde m  \nonumber \\
\partial_t x_0 &=&  V_m  \tilde m +\partial_y^2 x_0
\label{mv}
\end{eqnarray}
where $m_0=const $ is the steady-state mass of the soliton, and
$\tau =V(m_0) \partial_m h^-$. Seeking solution in the form $m,x_0
\sim \exp[\lambda t + i q y ] $, $q$ is the transverse modulation
wavenumber,  for the most unstable mode we obtain from Eq.
(\ref{mv}) the growthrate $\lambda$
\begin{equation}
\lambda = \frac{-q^2 (1+\zeta_2) -\tau + \sqrt{
(q^2(1-\zeta_2)-\tau)^2+4 V_m \zeta_1 q^2} }{2}\label{l0}
\end{equation}
Expanding Eq. (\ref{l0}) for $q \to 0$ we obtain $ \lambda \approx
\frac{1}{2}(2 V_m \zeta_1 /\tau -1) q^2 + O(q^4) $. The
instability occurs if $V_m \zeta_1 /\tau -1/2>0$. Substituting
$\tau$ and using $V_m/h_m= V_h$, we obtain a simple instability
criterion:
\begin{equation}
2 V_h \zeta_1 / V  > 1 \label{cond1}
\end{equation}
Eq. (\ref{cond1})  gives  a value of threshold $\alpha$ since
$\zeta_1 \sim \alpha$. For $\alpha<\alpha_c$ no instability
occurs, and the modulation wavelength diverges for $\alpha \to
\alpha_c$. Far away from the threshold we neglect $\tau$ and then
obtain for $\lambda(q)$:

\begin{eqnarray}
\lambda = |q| \sqrt {\zeta_1 V_m }-(1+\zeta_2)q^2/2 + O(q^3)
\label{l1}
\end{eqnarray}
The optimal wavenumber $q^*$  is given
\begin{equation}
q^* \sim \sqrt {\zeta_1 V_m } \sim \alpha \label{qm1}
\end{equation}
Fig. \ref{fig2} shows $\lambda(q) $ obtained by numerical
stability analysis of linearized Eqs. (\ref{conser}), (\ref{A1})
near the one-dimensional solution Eq. (\ref{soliton}). For
comparison is shown the solution to Eq. (\ref{l0}), with the
parameters extracted from the corresponding one-dimensional
steady-state problem Eqs. (\ref{conser1}),(\ref{A11}). One sees
that Eq. (\ref{l0}) gives correct description for small $q$,
however fails to predict $\lambda(q)$  in the whole range of $q$.
For this purpose one needs to include higher order terms. Thus,
Eq. (\ref{l0}) gives a correct description of the onset of
instability and qualitative estimate for the selected wavenumber
$q^*$. Inset to Fig. \ref{fig2} shows the dependence of optimal
wavenumber $q^*$ vs $\alpha$, obtained by numerical linear
stability analysis of the soliton solution. It shows almost linear
decrease of $q^*$ with $\alpha$ consistent with Eq. (\ref{qm1}).
For very small $\alpha$ the plot indicates that $q^* \to 0$ at
$\alpha \to \alpha_c$, consistent with Eq. (\ref{cond1}). From the
qualitative point of view, the transverse instability of planar
front is caused by the following mechanism:  local increase of
soliton mass results in the increase of its velocity and,
consequently, the ``bulging'' of the front. Due to the mass
conservation, the bulge depletes material in the neighboring areas
and further decreases their speed.

\begin{figure}[h]
\includegraphics[width=3.2in]{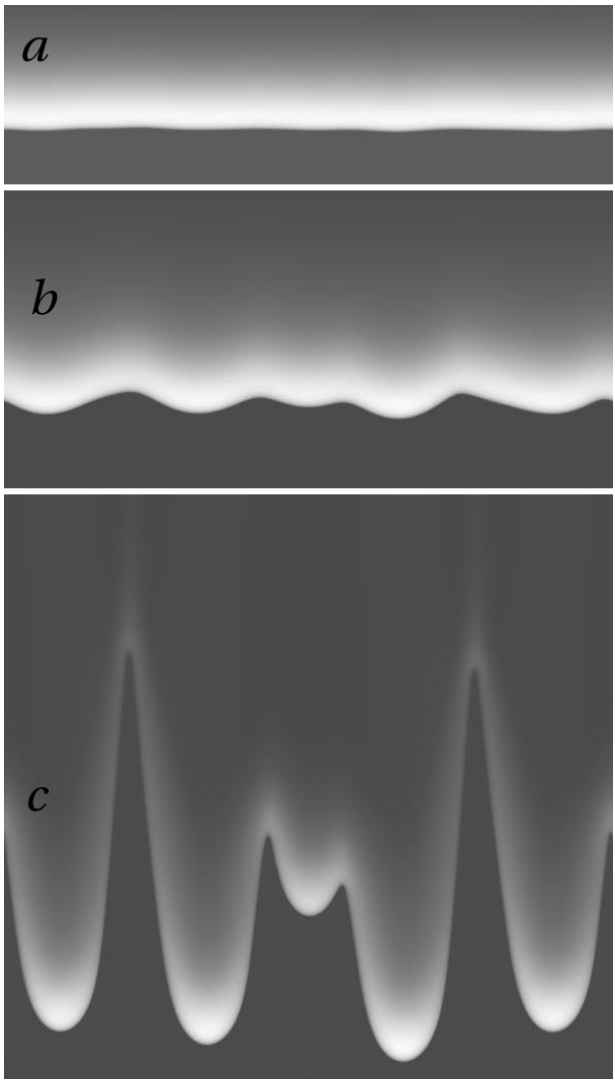}
\caption{Grey-coded images  of  $h(x,y)$ (white corresponds to
larger $h$) for  a) $t=170$, b) $t=300$ and c) $t=500$ units of
time. Domain size is 600 units in $x$  and 450 units in $y$
direction, only part of domain in $x$ direction is shown.
Parameters: $\delta=1.16$, $\alpha=0.14$, $\beta=2$ and initial
height $h_0=2.285$.  } \label{fig3}
\end{figure}

To study   the evolution of the avalanche front beyond the initial
linear instability regime, a  fully two-dimensional numerical
analysis of Eqs. (\ref{conser}), (\ref{A1}) was performed.
 Integration was performed in a rectangular domain with periodic boundary conditions in $x$
and $y$ directions. The number of mesh points was up to
$1200\times600$ or higher. As an initial condition we used a flat
state $h=h_0$ with a narrow stipe $h=h_0+2$ deposited along the
$y$-direction. To trigger the transverse instability, small noise
was added to the initial conditions. The initial conditions
rapidly developed into a quasi-one-dimensional solution  described
by Eq. (\ref{soliton}). Due to the periodicity in the
$x$-direction, the soliton could pass through the integration
domain several times. It allowed us to perform analysis in a
relatively small domain in the $x$-direction. The transverse
modulation of the soliton leading front was observed after about
100 units of time for the parameters of Fig. \ref{fig3}. We
observe that modulation initially grows in amplitude, eventually
coarsens and leads to the formation of large-scale finger
structures.

At the qualitative level the agreement between theory
 and  experimental results of
Mallogi et al. \cite{Malloggi:2005} is impressive. (i) Existence
 of steady-state soliton-like avalanches propagating
downhill with a shape similar to experiment. (ii) Generic zero
wave number (longwave) transverse instability compatible with the
experimental divergence of the selected wavelength close to the
instability threshold. Far from the threshold, linear growth rate
dependence with $q$  compatible with measurements. (iii)
Coarsening in the later development of the instability. (iv)
Fingering instability with localized droplet-like avalanches (also
similar to those described in \cite{Borzsonyi:2005}). The analysis
predicts that the transverse instability ceases to exist when the
rescaled transport coefficient $\alpha$ decreases (see Fig.
\ref{fig2}). In the present form, the model does not provide an
explicit relation between $\alpha$ and the chute angle $\varphi$
(since $\alpha$ depends also on $\tau_\rho$). Nevertheless,
molecular dynamics studies indicate that the OP diffusion
coefficient $D_\rho=l_\rho^2/\tau_\rho$ increases with pressure
\cite{VTA03}. Since the pressure is proportional to the sediment
height $h_{0}$ which increases as the angle $\varphi$ decreases,
it results in the decrease of $\tau_\rho$. Thus, with the decrease
of angle $\bar \varphi$ the instability should disappear, in
agreement with experiment where the soliton is found stable at
lower inclination angles.

An important question remains is  how to bring to a more
quantitative level the comparison between theory  and the
experimental measurements. In this perspective, a challenging
question
 is to deeply understand  the qualitative differences between smooth glass bead and rough sandy materials as far as the effective
 flow rules and avalanche shapes are concerned. This work calls for more systematic measurements centered on the soliton velocity
 dependence with the flowing mass for various materials and the possible identification of an instability threshold for glass beads.
 Such results would allow a more precise assessment of the model parameters and could lead the way to a reliable and predictable
 modelling of granular avalanches.
 The  fingering patterns
bear remarkable similarities with those existing in thin films
flowing down inclined surfaces, both with clear and particle-laden
fluids \cite{Troian:1989}.  However, the physical mechanisms
leading to this fingering  are likely dissimilar: in fluid films,
it is driven (and stabilized) by the surface tension, whereas in
the  granular flow case, the  surface tension plays no role. We
thank Olivier Pouliquen, Bruno Andreotti, Stephane Douady, Lev
Tsimring, Tamas B\"orzs\"onyi and Robert Ecke for discussions and
help. IA was supported by US DOE, Office of Science, contract
W-31-109-ENG-38.

\end{document}